\documentclass{aa}
\usepackage{graphicx}
\usepackage{natbib}
\usepackage{txfonts}
\usepackage{amssymb}
\bibpunct{(}{)}{;}{a}{}{,} 
\begin{document}

   \title{Network oscillations at the boundary of an equatorial coronal hole}
   \author{H. Tian\inst{1,2}
           \and
           L.-D. Xia\inst{3}
          }
   \offprints{H. Tian}
   \institute{Max-Planck-Institut f\"ur Sonnensystemforschung, Katlenburg-Lindau, Germany\\
             \email{tian@mps.mpg.de}
             \and
             Department of Geophysics, Peking University, Beijing, China
             \and
             School of Space Science and Physics, Shandong Univ. at Weihai, Weihai, Shandong, China\\
             \email{xld@sdu.edu.cn}
             }
   \date{}

\abstract
{}
{We investigate intensity oscillations observed simultaneously in
the quiet chromosphere and in the corona, above an enhanced network
area at the boundary of an equatorial coronal hole.}
{A Fourier analysis is applied to a sequence of images observed in
the 171~{\AA} and 1600~{\AA} passbands of TRACE. Four interesting
features above the magnetic network are further investigated by
using a wavelet analysis. }
{Our results reveal that, in both the 171~{\AA} and 1600~{\AA}
passbands, oscillations above the magnetic network show a lack of
power at high frequencies (5.0-8.3 mHz), and a significant power at
low (1.3-2.0 mHz) and intermediate frequencies (2.6-4.0 mHz). The
global 5-min oscillation is clearly present in the 4 analyzed
features when seen in the 1600~{\AA} passband, and is also found
with enhanced power in feature 1 (leg of a large coronal loop) and
feature 2 (legs of a coronal bright point loop) when seen in the
171~{\AA} passband. Two features above an enhanced network element
(feature 3 and feature 4) show repeated propagating behaviors with a
dominant period of 10 minutes and 5 minutes, respectively. }
{We suggest these oscillations are likely to be slow
magneto-acoustic waves propagating along inclined magnetic field
lines, from the lower solar atmosphere into the corona. The energy
flux carried by these waves is estimated of the order of $\rm40~erg~
cm^{-2}~s^{-1}$ for the 171~{\AA} passband and is far lower than the
energy required to heat the quiet corona. For the 1600~{\AA}
passband, the energy flux is about $\rm1.4\times10^6~erg~
cm^{-2}~s^{-1}$, which is about one third of the required energy
budget for the chromosphere.}


\titlerunning{Network oscillations at the boundary of an equatorial coronal hole}
\authorrunning{H. Tian and L.-D. Xia}
\maketitle

\section{Introduction}

The solar atmosphere is highly structured by the magnetic field. The
chromospheric network, which is the upward extension of the
supergranular boundaries above the photosphere, is characterized by
clusters of magnetic flux concentrations \citep{Gabriel1976}. Part
of the network flux expands with height
\citep{PatsourakosEtal1999,TianEtal2008a} and opens into the corona
with a funnel shape, while the rest of the network consists of a
dense population of low-lying loops with lengths less than $10^4$~km
\citep{DowdyEtal1986,Peter2001}.

The observation of oscillations in the chromosphere and corona can
help us understand the magnetic structure of the solar atmosphere,
and provides valuable insight into the unresolved coronal heating
problem.

It has been shown that oscillations with different frequencies (0.6
mHz-2.7 mHz) are present in the chromospheric network
\citep{CurdtHeinzel1998, LitesEtal1993, CurdtHeinzel1998,
CauzziEtal2000}. \cite{McAteerEtal2002} found waves with a period of
4-15 minutes (1-4 mHz) in the central portions of network bright
points, and suggest that these waves are possibly magneto-acoustic
or magneto-gravity modes. More recently, \cite{McAteerEtal2004}
found that the most frequent network oscillation has a period of 283
s, with a lifetime of 2-3 cycles in four TRACE (Transition Region
and Coronal Explorer) passbands centered at 1700, 1600, 1216, and
1550~{\AA}. The power of the intensity fluctuations in the
1600~{\AA} passband (UV continuum at $4-10\times10^{3}~$K plus
Fe~{\sc{ii}} at $1.3\times10^{4}~$K), which is formed at the
temperature minimum \citep{McAteerEtal2004}, is found to peak at 3-5
mHz and considered as a mixture of photospheric 5-min oscillations
and chromospheric 3-min oscillations \citep{FossumCarlsson2005}. The
so-called $''$magnetic shadows$''$, which surround magnetic network
elements and show a lack of oscillatory power in the period range of
2-3 minutes, have received much attention
\citep{McIntoshJudge2001,KrijgerEtal2001,VecchioEtal2007} and have
been considered to play an important role in the trapping of
high-frequency magneto-acoustic oscillations
\citep{SrivastavaEtal2008}.

Oscillations with different periods have also been found in
different regions above the chromosphere. In the active region,
3-min oscillations have been found in the transition region and
corona in sunspot regions \citep{Fludra1999, BrynildsenEtal1999,
DeMoortelEtal2002}, while 5-min oscillations are more likely to be
found in $^{\prime\prime}$non-sunspot$^{\prime\prime}$ loops
\citep{DeMoortelEtal2002} and above AR plage
\citep{DePontieuEtal2003,DePontieuEtal2005}. These oscillations are
suggested to be due to upward-propagating slow magneto-acoustic
waves. The 5-min oscillation is likely to be guided along inclined
magnetic flux tubes, which can decrease the acoustic cutoff
frequency to allow the low-frequency photospheric oscillations
propagate into the outer atmosphere
\citep{DePontieuEtal2004,McIntoshJefferies2006,HansteenEtal2006}. In
the open-field regions,  quasi-periodic perturbations with periods
of 10-15 minutes \citep{DeForestGurman1998} and 10-25 minutes
\citep{BanerjeeEtal2000} have been found in polar plumes. These
low-frequency perturbations are explained as compressive waves (such
as sound waves or slow-mode magneto-acoustic waves) propagating
along the plumes \citep{OfmanEtal1999}. In spicules, the 5-min
oscillation is also present \citep{XiaEtal2005}, and might be driven
by p-modes \citep{DePontieuEtal2004}. By measuring phase delays
between intensity oscillations and between velocity oscillations of
different line pairs, \cite{OSheaEtal2007} concluded that
propagating slow magneto-acoustic waves are present in coronal holes
and that they occur preferentially in bright regions that are
associated with magnetic field concentrations in the form of loops
or bright points. Periodic oscillations (7-64 minutes) in coronal
bright points have also been found by \cite{UgarteEtal2004a},
\cite{UgarteEtal2004b}, and \cite{TianEtal2008b}. By using the new
\emph{XRT} data, \cite{KrijgerEtal2008} recently found that X-Ray
bright point emission shows several significant peaks at different
frequencies corresponding to time scales that range from a few
minutes to hours.

In this contribution, we investigate intensity oscillations observed
simultaneously in the quiet chromosphere and in the corona, above an
enhanced network area at the boundary of an equatorial coronal hole.

\section{Observations and data analysis}

The data set analyzed here contains a sequence of images observed
from 07:40 to 08:20 UTC on November 08 in 1999 in two passbands
(1600~{\AA} and 171~{\AA}) of the TRACE instrument. The emissions of
the 1600~{\AA} and 171~{\AA} passbands are formed in the
chromosphere (at temperature minimum) and in the lower corona,
respectively. The sequence of 171~{\AA} only lasted from 07:40 to
08:12. The pixel size is $0.5^{\prime\prime}$ for the 1600~{\AA}
images, and $1.0^{\prime\prime}$ for the 171~{\AA} images. Standard
software for calibrating and correcting the TRACE data was applied
to this data set, including removal of cosmic rays, subtraction of
the dark current, normalization of the counts, and so on. We used
the cross correlation technique to do the coalignment of the images
in the sequence.

\begin{figure}
\resizebox{\hsize}{!}{\includegraphics{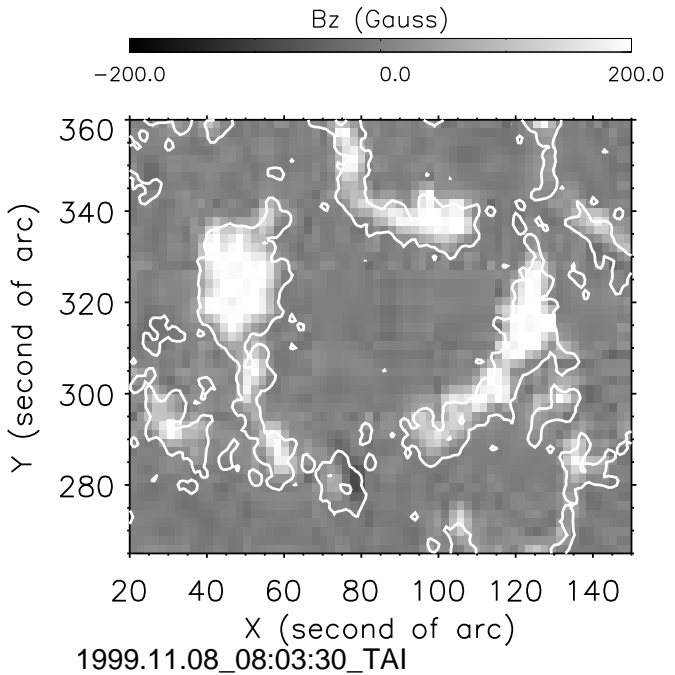}} \caption{MDI
magnetogram of the studied region taken at 08:03:30 UTC on November
08 in 1999. The white contours correspond to bright lanes seen in
the averaged image of the 1600~{\AA} passband.} \label{fig.1}
\end{figure}

\begin{figure*}
\resizebox{\hsize}{!}{\includegraphics{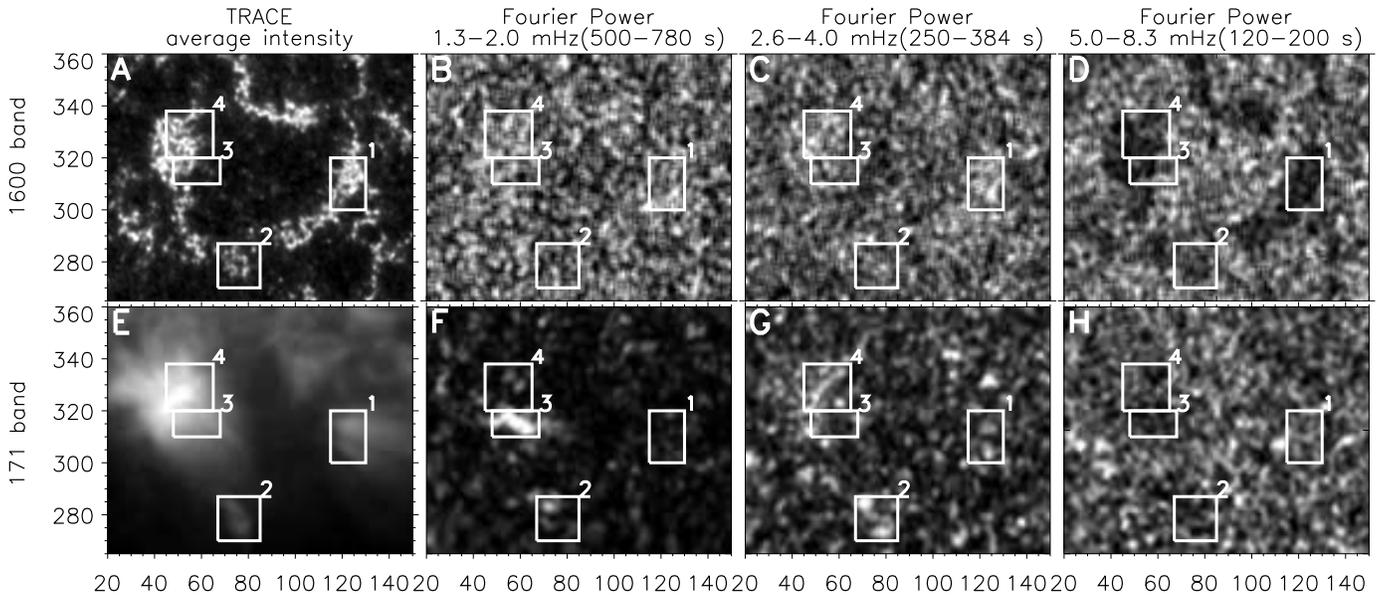}}
\caption{~Panel A-D: The average intensity image of the 1600~{\AA}
passband, and maps of Fourier power in frequency ranges of 1.3-2.0
mHz, 2.6-4.0 mHz, and 5.0-8.3 mHz. ~Panel E-H: similar to Panel A-D,
but for the 171~{\AA} passband.} \label{fig.2}
\end{figure*}

We extracted a sequence of sub-images with a size of
$130^{\prime\prime}\times95^{\prime\prime}$ from the original data
set. The sub-images were taken around the boundary of an equatorial
coronal hole above an enhanced network area. The average intensity
images of the studied region in the 1600~{\AA} and 171~{\AA}
passbands can be found in Fig.~\ref{fig.2}A and E, respectively. The
chromospheric network pattern is clearly seen in the 1600~{\AA}
passband. The coronal hole and the surrounding quiet Sun can be
easily discerned in the image of the 171~{\AA} passband.
Fig.~\ref{fig.1} shows the magnetogram corresponding to the studied
region.

The cadence of our TRACE data was about 30~s. At a few moments there
were irregular gaps. This problem was overcome by applying a linear
interpolation with the help of the \emph{INTERPOL} function in the
IDL software. We then created a sequence of running difference
images, by subtracting each image from the image taken 30~s earlier.
This sequence of running difference images was further analyzed to
search for periodic signatures.

We applied a standard Fourier analysis to the sequence of running
difference images of each passband. At each spatial pixel, we
obtained a frequency-power curve. Then we added the total power
respectively in three different frequency ranges, and calculated the
percentage of this value relative to the total power in the full
frequency range. In this way we obtained a relative power map in
each frequency range. The three frequency ranges are 1.3-2.0 mHz
(500-780 s, low frequencies), 2.6-4.0 mHz (250-384 s, intermediate
frequencies), and 5.0-8.3 mHz (120-200 s, high frequencies). The
power maps can be found in Fig.~\ref{fig.2}.

The intensity images and maps of Fourier power reveal several
interesting features. We chose four features for a more detailed
analysis. They are shown in Fig.~\ref{fig.2} and outlined in white.
By checking the original TRACE images and the full-disk MDI
magnetogram, we found that feature 1 might be a leg of a large
coronal loop, feature 2 is a coronal bright point. Feature 3 and
feature 4 are above an enhanced network element and their enhanced
power forms an elongated shape at low and intermediate frequencies,
respectively.

\begin{figure*}
\resizebox{\hsize}{!}{\includegraphics{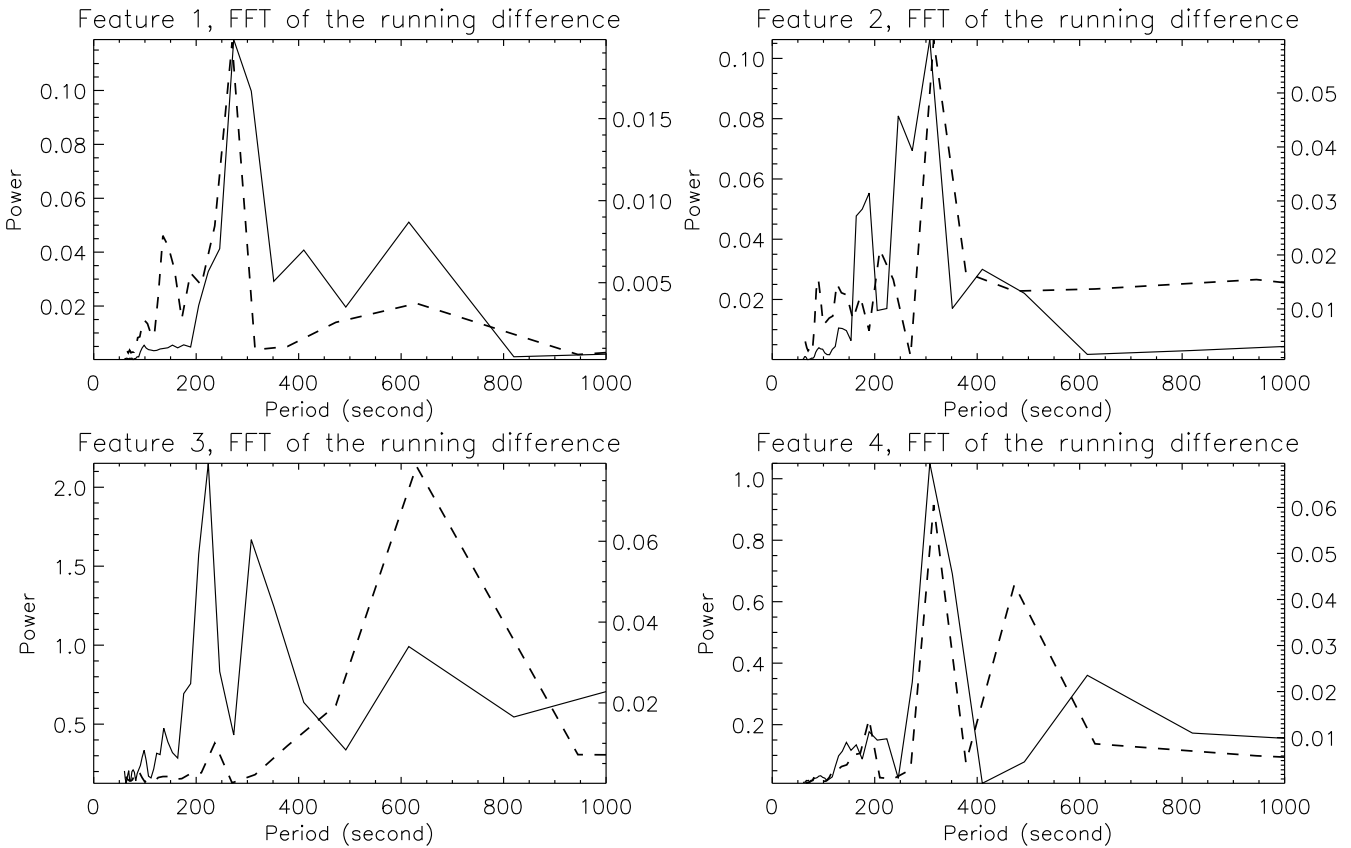}} \caption{~The
Fourier spectra for the four features. The solid and dashed curves
correspond to the power of the 1600~{\AA} and 171~{\AA} passbands
and their scales are given on the left-hand and right-hand y-axes,
respectively.} \label{fig.3}
\end{figure*}

\begin{figure*}[tp]
\centering
\begin{minipage}[t]{0.45\textwidth}
{\includegraphics[width=\textwidth]{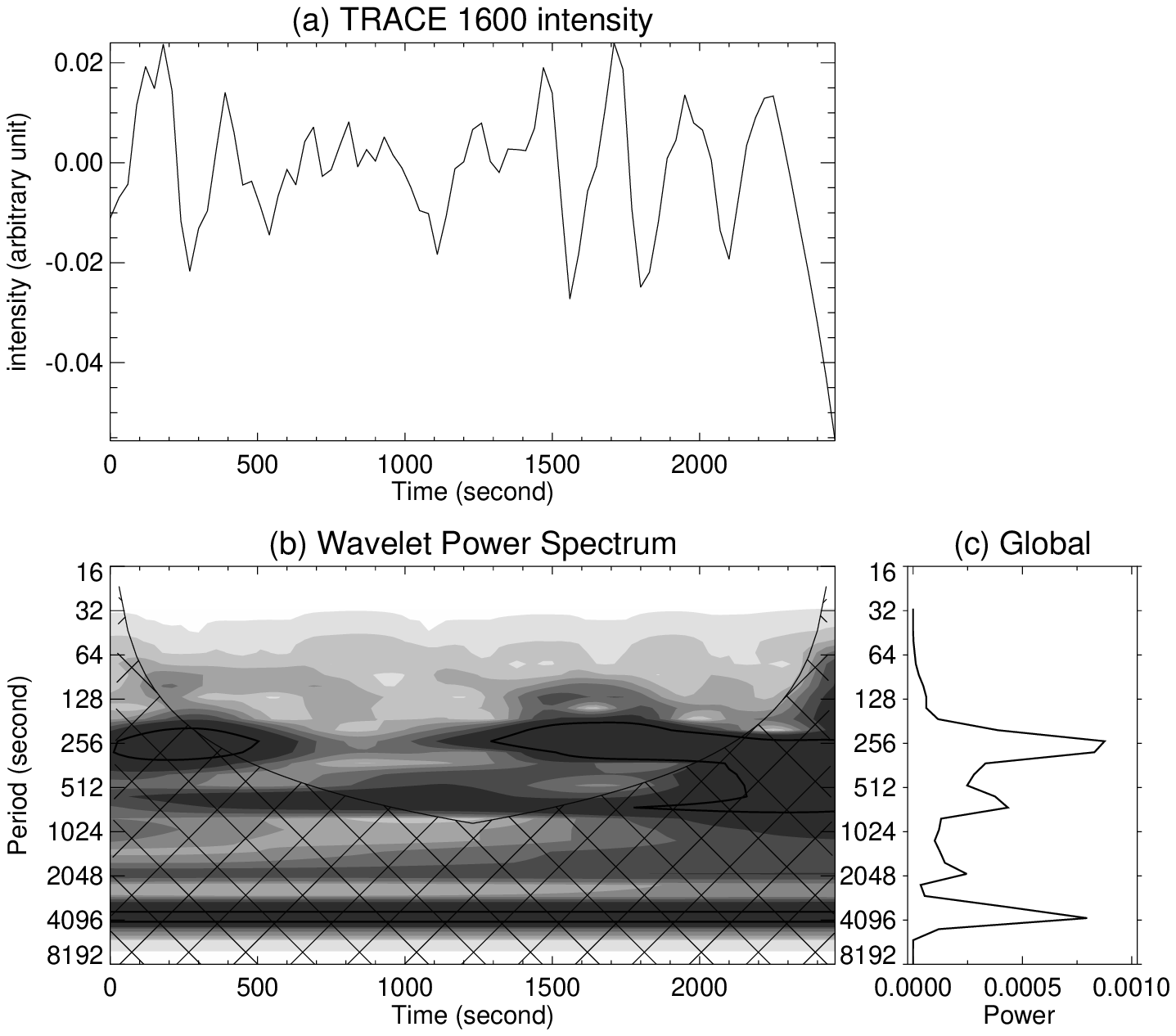}}
\end{minipage}
\begin{minipage}[t]{0.45\textwidth}
{\includegraphics[width=\textwidth]{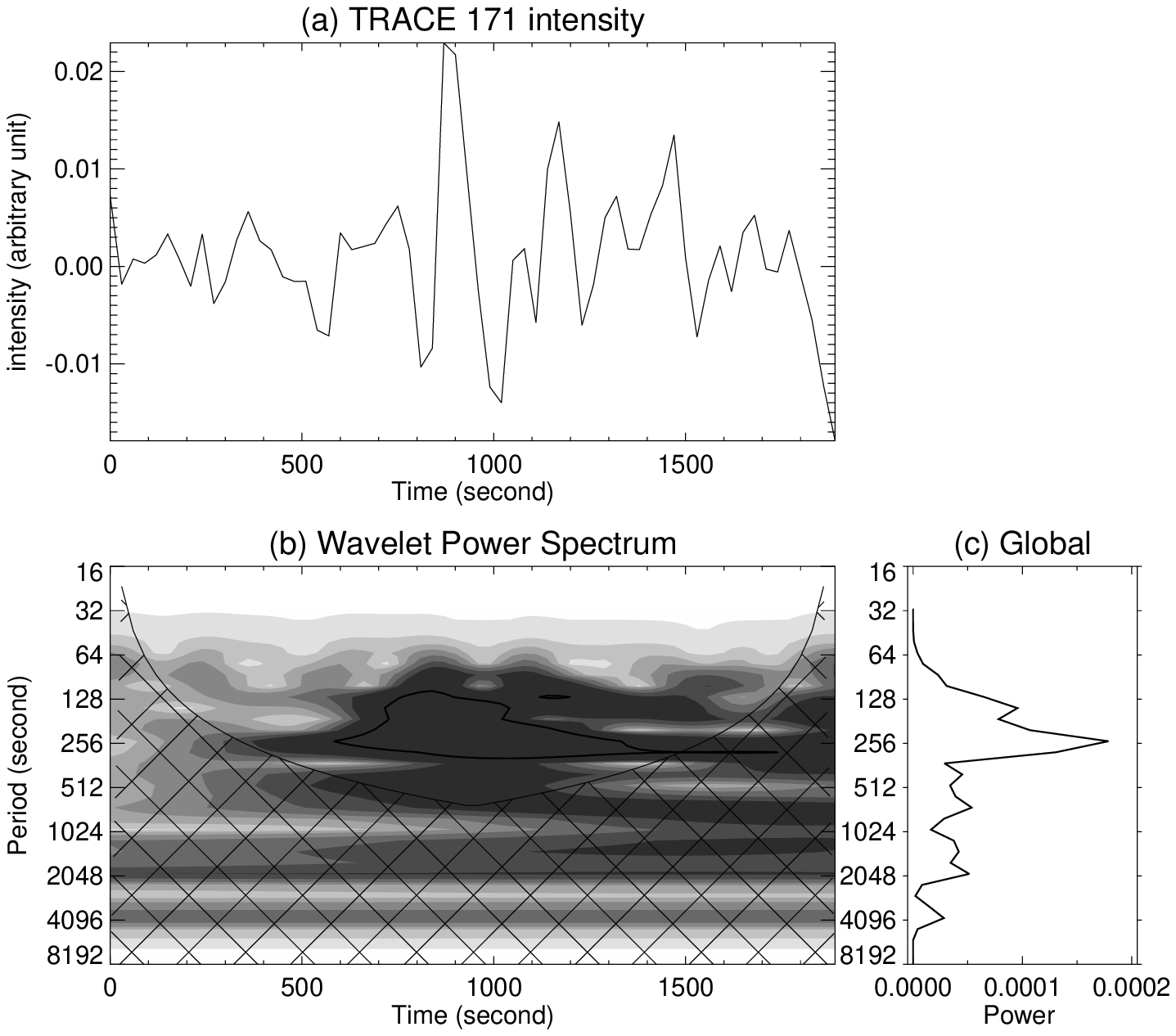}}
\end{minipage}
\begin{minipage}[b]{\textwidth}
\caption{~Wavelet power spectra of the 1600~{\AA} (left) and
171~{\AA} (right) passbands for feature 1. (a) The running
difference light curve. (b) Time/period variation of the wavelet
power spectrum. Cross-hatched regions indicate the
$^{\prime\prime}$cone of influence$^{\prime\prime}$. The darker
parts represent higher power, and the contours correspond to the
95\% confidence level. (c) Global wavelet. } \label{fig.4}
\end{minipage}
\end{figure*}

\begin{figure*}[tp]
\centering
\begin{minipage}[t]{0.45\textwidth}
{\includegraphics[width=\textwidth]{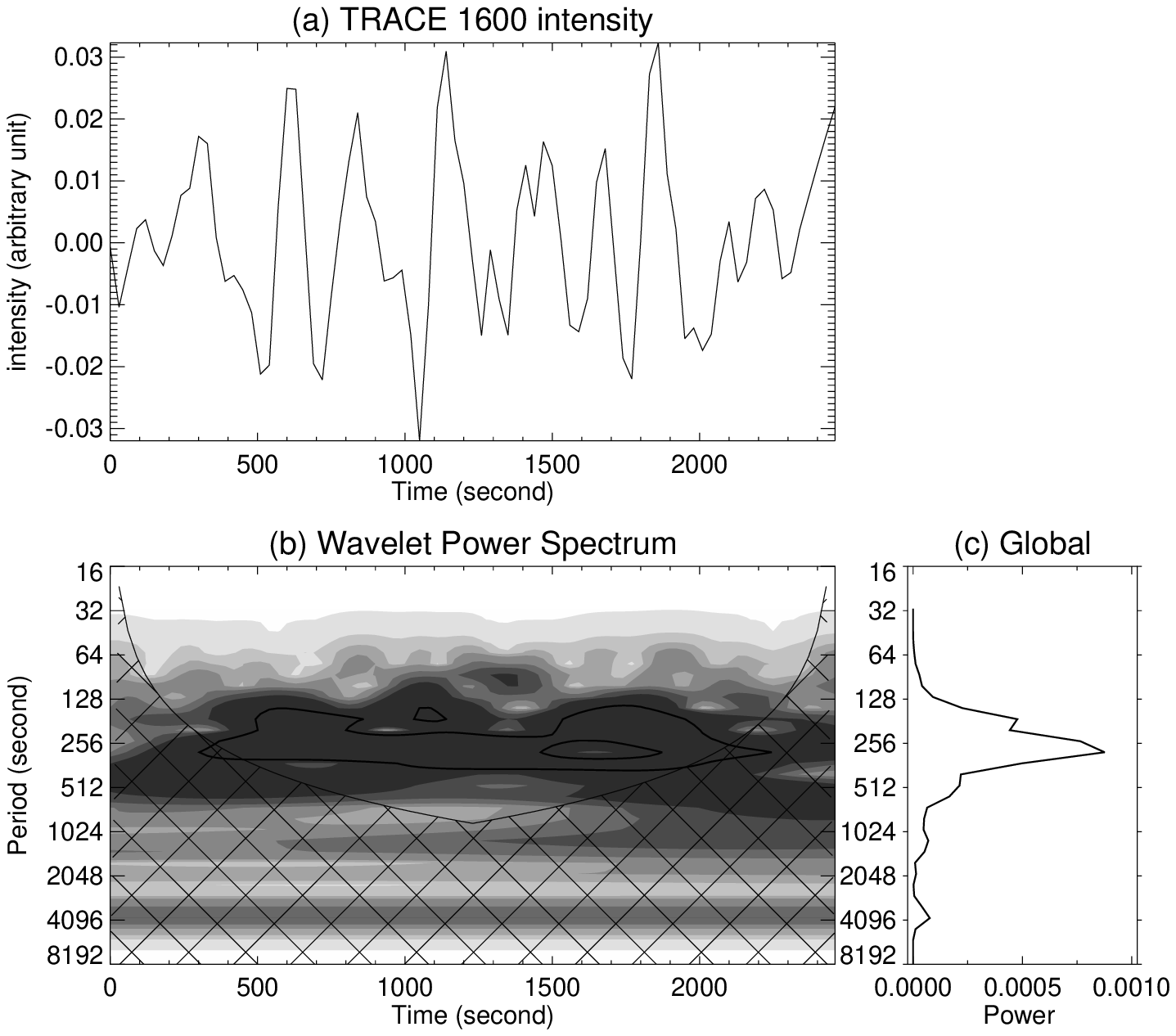}}
\end{minipage}
\begin{minipage}[t]{0.45\textwidth}
{\includegraphics[width=\textwidth]{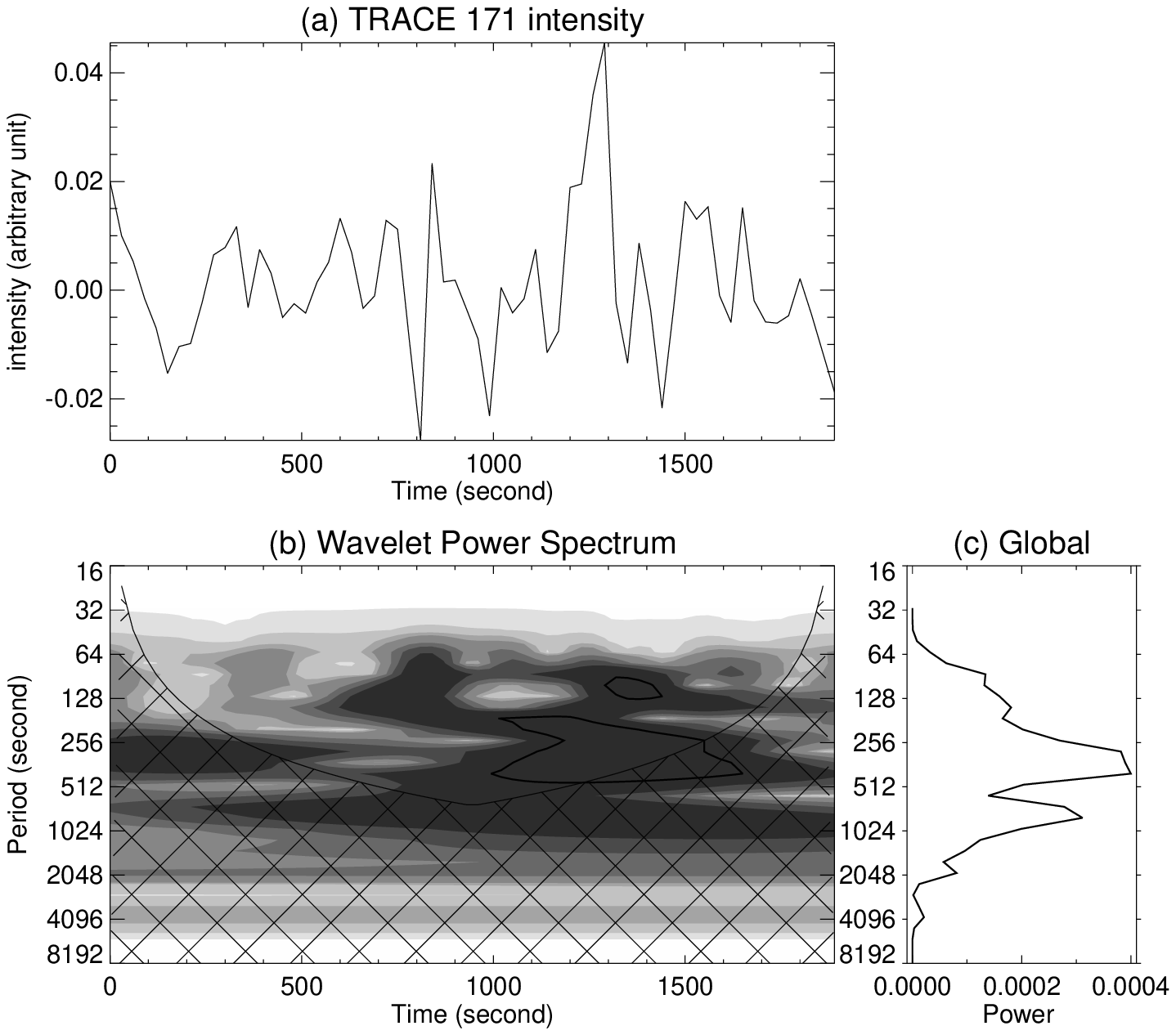}}
\end{minipage}
\begin{minipage}[b]{\textwidth}
\caption{Same as Fig.~\ref{fig.4} but for feature 2. } \label{fig.5}
\end{minipage}
\end{figure*}

\begin{figure*}[tp]
\centering
\begin{minipage}[t]{0.45\textwidth}
{\includegraphics[width=\textwidth]{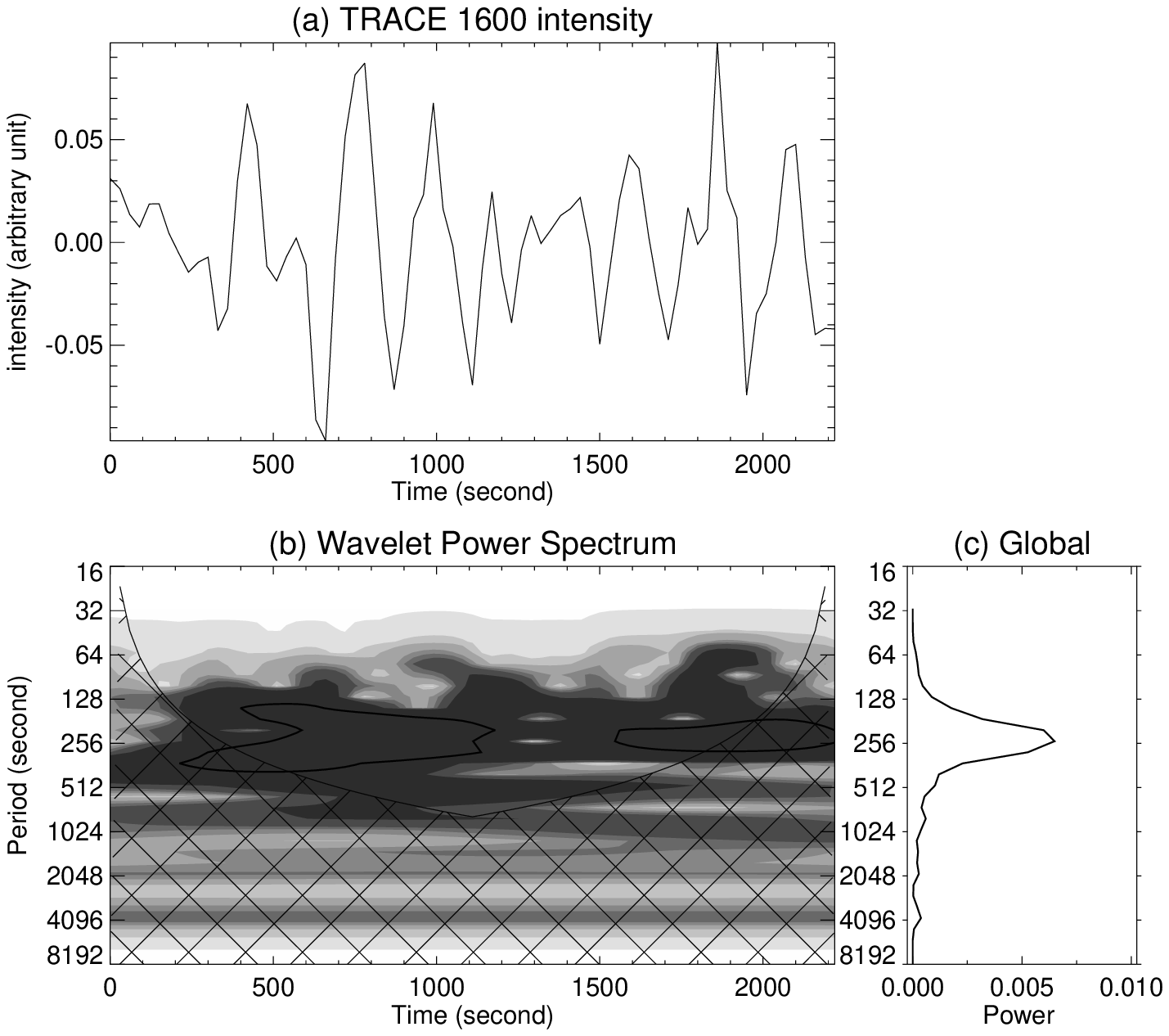}}
\end{minipage}
\begin{minipage}[t]{0.45\textwidth}
{\includegraphics[width=\textwidth]{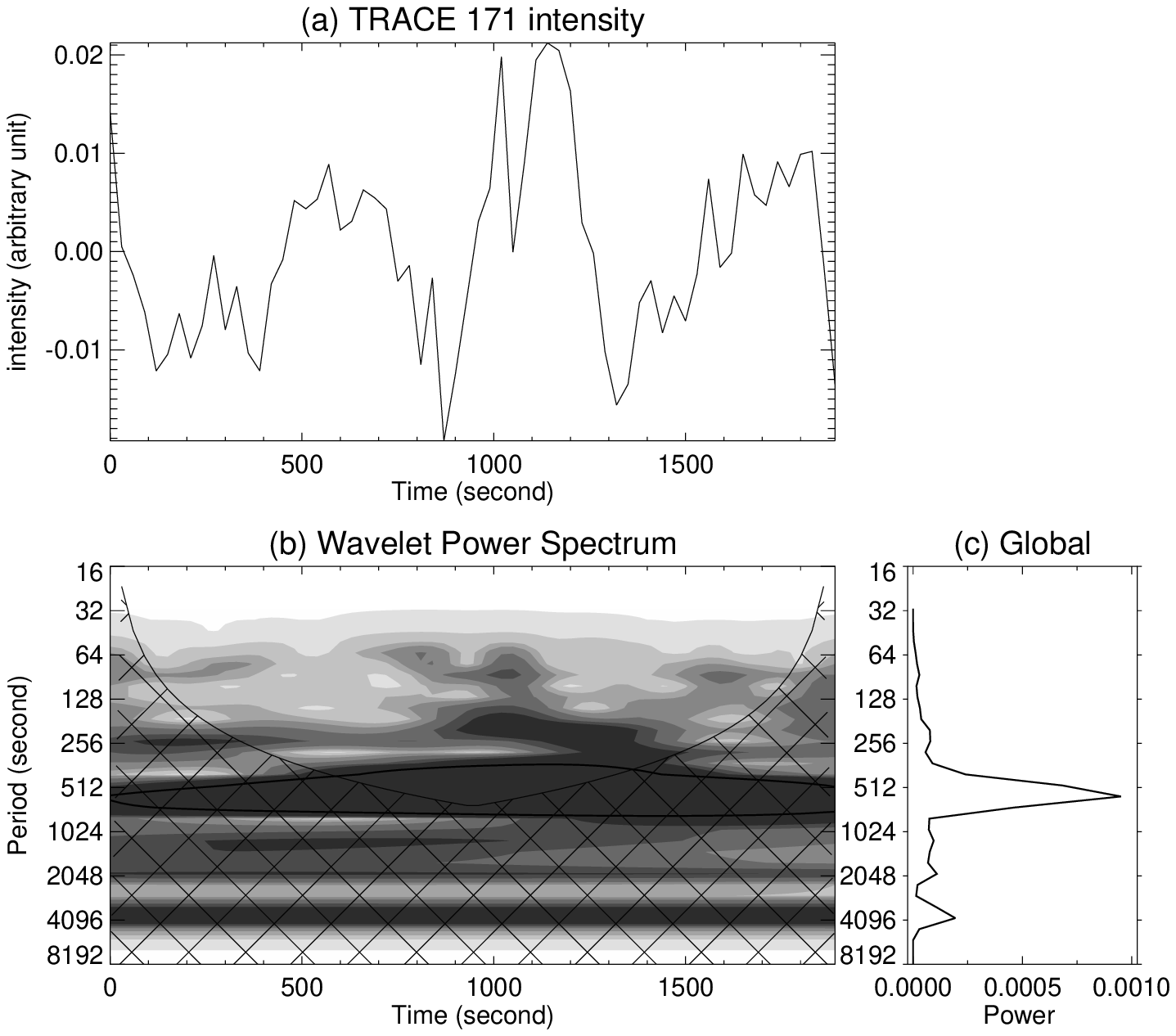}}
\end{minipage}
\begin{minipage}[b]{\textwidth}
\caption{Same as Fig.~\ref{fig.4} but for feature 3. } \label{fig.6}
\end{minipage}
\end{figure*}

\begin{figure*}[tp]
\centering
\begin{minipage}[t]{0.45\textwidth}
{\includegraphics[width=\textwidth]{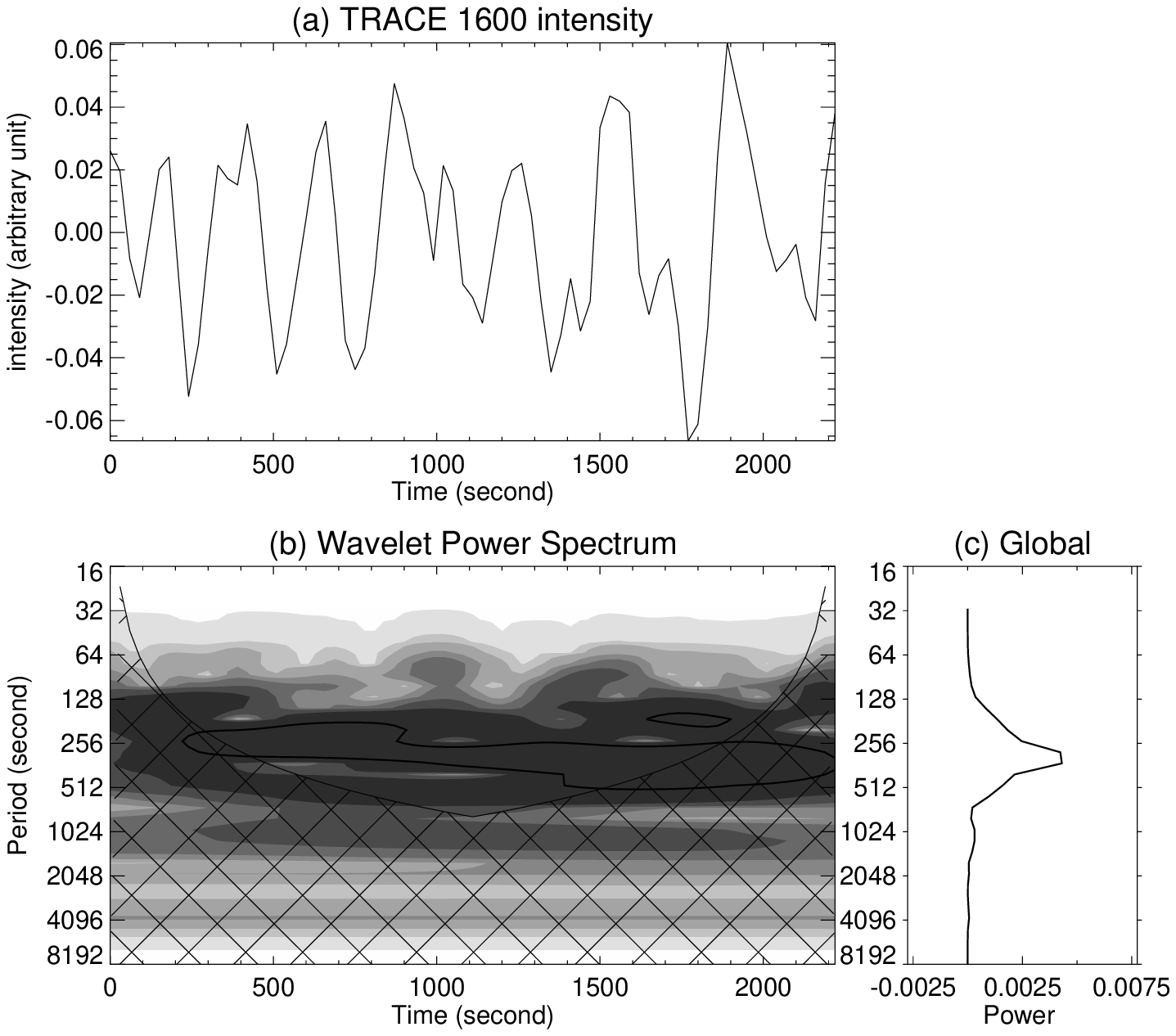}}
\end{minipage}
\begin{minipage}[t]{0.45\textwidth}
{\includegraphics[width=\textwidth]{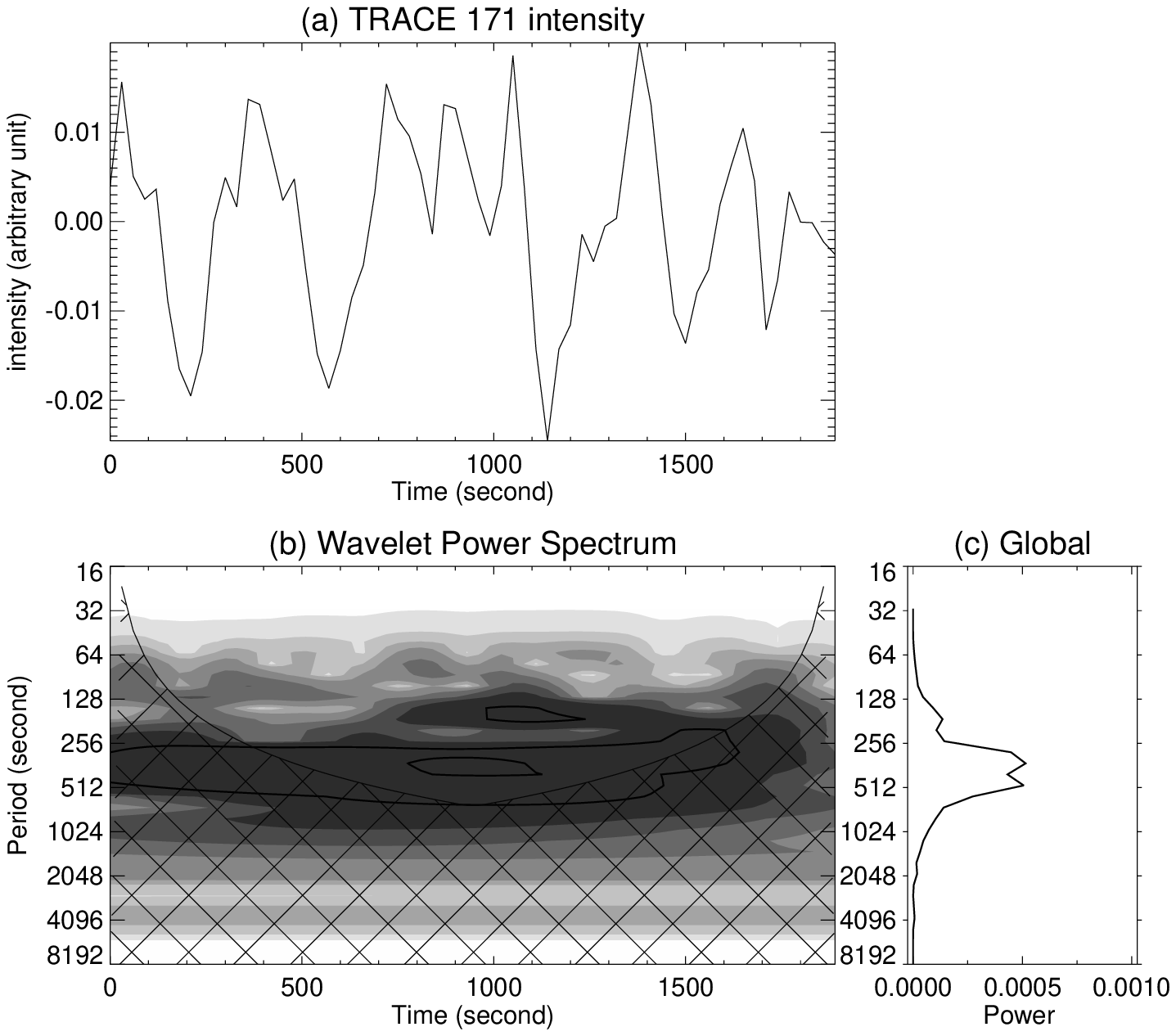}}
\end{minipage}
\begin{minipage}[b]{\textwidth}
\caption{Same as Fig.~\ref{fig.4} but for feature 4. } \label{fig.7}
\end{minipage}
\end{figure*}

We averaged the intensity in the outlined rectangular area for each
feature, and created the corresponding running difference light
curve. These curves are shown in the upper panels of
Fig.~\ref{fig.4} to Fig.~\ref{fig.7}. Then we applied a Fourier
analysis and a wavelet analysis to each running difference light
curve. The Fourier spectra are shown in Fig.~\ref{fig.3}. The solid
and dashed curves correspond to the power of the 1600~{\AA} and
171~{\AA} passband, respectively.

The localized nature of the wavelet transform allows us to study the
duration of any statistically significant oscillations as well as
their period \citep{BanerjeeEtal2000}. By decomposing a time series
into time-frequency space, one is able to determine both the
dominant modes of variability and how those modes vary in time
\citep{TorrenceCompo1998}. Here we also performed a wavelet analysis
for the running difference light curves of the four features, in
order to find the most reliable periods. We chose the Morlet wavelet
function, defined as a sine wave modulated by a Gaussian window, for
our analysis. The wavelet transform suffers from edge effect at both
ends of the time series. And this effect is important in regions
defined as the $^{\prime\prime}$cone of influence$^{\prime\prime}$
(COI). To check whether the periodic signatures present in the
wavelet spectrum are real or not, we have to perform a significance
test. Here we chose a confidence level of 95\%. The description of
the wavelet function, COI, and significance test can be found in
\cite{TorrenceCompo1998}.

The wavelet power spectra of the four features are shown in
Fig.~\ref{fig.4} to Fig.~\ref{fig.7}. Cross-hatched regions indicate
the $^{\prime\prime}$cone of influence$^{\prime\prime}$. The darker
parts represent higher power, and the contours correspond to the
95\% confidence level. We also averaged each wavelet power spectrum
over time and plotted the global wavelet spectrum.

\section{Results and discussion}

The main motivation of this work is to search for the signature of
oscillations at the boundary of an equatorial coronal hole, and to
study oscillations observed almost simultaneously in the
chromosphere and corona above the magnetic network. The Fourier and
wavelet analyses show that obvious oscillations really exist in our
data.

Fig.~\ref{fig.2}B-D show the maps of Fourier power in different
frequency ranges in the 1600~{\AA} passband. It is very clear that
the network oscillations reveal a lack of power at high frequencies,
and a significant power at low and intermediate frequencies. The
reason why the power at high frequencies in the internetwork region
is stronger than that in the network lane might be a result of the
network magnetic structure. It is known that part of the network
flux expands with height and opens into the
corona\citep{DowdyEtal1986,Peter2001,PatsourakosEtal1999,TianEtal2008a}.
Thus, photospheric oscillations with frequencies above the acoustic
cutoff (about 5.5 mHz) can propagate into the chromosphere, where
the network fluxes begin to stretch into the region above the
internetwork in the shape of canopies \citep{SrivastavaEtal2008}.
Below the canopy field, magnetic shadows with reduced oscillatory
power at high frequencies are present within and immediately around
the magnetic network \citep{VecchioEtal2007}. But the significant
power (seen in Ca~{\sc{ii}} 854.2 nm) directly above the magnetic
network in \cite{VecchioEtal2007} is not very clear in our data set,
which might be due to the different formation height between
Ca~{\sc{ii}} and the 1600~{\AA} passband of TRACE, or due to the
relatively low spatial resolution of our data set.

The result of a significant power at low and intermediate
frequencies might be explained as a leakage of low-frequency
photospheric oscillations into the chromosphere through
$^{\prime\prime}$magneto-acoustic portals$^{\prime\prime}$ which are
positioned within the network. It might provide a significant energy
source to heat the quiet chromosphere
\citep{JefferiesEtal2006,VecchioEtal2007}.

As for the power maps of the 171~{\AA} passband, we can make a
similar conclusion to that of the 1600~{\AA} passband. But the
significant powers at low and intermediate frequencies are more
constricted and isolated in the network lane. Also, the regions with
reduced power at high frequencies are not so prominent and are more
diffused. It might be a natural result of the diffusion and
expansion of the magnetic structure from the chromosphere to the
corona.

As we stated above, feature 1 corresponds to a leg of a large
coronal loop system. Fig.~\ref{fig.2}C and Fig.~\ref{fig.2}G reveal
a significant power at intermediate frequencies in this loop leg,
both in the 1600~{\AA} and 171~{\AA} passbands. The period
corresponding to these frequencies is around 5 minutes. Although
oscillations with periods 180-420~s have already been reported by
\cite{DeMoortelEtal2000}, it is still very interesting to note that
in feature 1 the two bright dots seen in Fig.~\ref{fig.2}G
correspond exactly to the two parts of the loop leg seen in
Fig.~\ref{fig.2}E. This result indicates that in a loop system, the
strongest power of the 5-min oscillation locates at the lower part
of the leg. As discussed in \cite{DeMoortelEtal2000}, the 5-min
oscillation might be due to the propagating slow magneto-acoustic
waves. From Fig.~\ref{fig.3} and Fig.~\ref{fig.4}, we find that the
power peak around 5 minutes is clearly present at most of the time.
There is a weak peak at 10 minutes for the Fourier spectra in both
the 1600~{\AA} and 171~{\AA} passbands, but not significant in the
wavelet power spectra.

Feature 2 corresponds to a coronal bright point, which is
characterized by an enhanced coronal emission associated with
bipolar magnetic field. \cite{UgarteEtal2004a} and
\cite{UgarteEtal2004b} studied periodic oscillations of coronal
bright points, and found oscillations with a period of 400-1100~s.
Longer period (8-64 minutes) oscillations were also detected by
\cite{TianEtal2008b}. Here from Fig.~\ref{fig.3} and
Fig.~\ref{fig.5} we find that the oscillatory power of this bright
point has a major peak around 5 minutes, and a second peak around 3
minutes. If we compare Fig.~\ref{fig.2}E and Fig.~\ref{fig.2}G, we
can find that the strongest power of feature 2 corresponds to the
two parts of the bright emission. These two parts might be the two
legs of the loop system associated with the bright point.

Feature 3 locates above the outer part of an enhanced network
element. The most interesting point is that the enhanced Fourier
power at low frequencies has an obvious elongated shape in the
171~{\AA} passband. The movie of the image sequence clearly reveals
that a periodic motion is present in this region, and the direction
of this motion is exactly along this elongated shape. The Fourier
and wavelet analyses both reveal a 10-min oscillation in the
171~{\AA} passband. This oscillation is present in the entire
duration of the time series, which can be seen in Fig.~\ref{fig.6}.
While in the 1600~{\AA} passband, the Fourier spectrum shows three
peaks at 3, 5, and 10 minutes. However, only the 3-min and 5-min
oscillations are considered as significant in the wavelet power
spectrum. It has been known that strongly inclined magnetic fields
can significantly decrease the acoustic cut-off frequency
\citep{DePontieuEtal2004,McIntoshJefferies2006,HansteenEtal2006},
and thus allow high-frequency photospheric oscillations (above the
acoustic cut-off frequency 5.5 mHz) to propagate into the
chromosphere and corona. Fig.~\ref{fig.2}E gives us the impression
that the magnetic field of feature 3 seems to be strongly inclined,
and thus is very likely to guide the high-frequency oscillations
from the lower part of the solar atmosphere to the corona. The weak
peak at 10 minutes in the Fourier spectrum of the 1600~{\AA}
passband, although not prominent in the wavelet spectrum, might
still be the counterpart of the dominant 10-min oscillation seen in
the 171~{\AA} passband.

As for feature 4, which is above the same enhanced network element
as feature 3 but in a different location, the power peak around 5
minutes is also clear in both the Fourier and the wavelet spectra
for each bandpass. From Fig.~\ref{fig.7} we can see that in the
1600~{\AA} passband the period is initially around 4 minutes, then
increases to about 6 minutes in the second half of the duration. In
the 171~{\AA} passband, a period of 5-8 minutes is present through
almost the entire duration. The strong power of the 5-min
oscillation also seems to form an elongated shape in the 171~{\AA}
passband shown in Fig.~\ref{fig.2}G. A periodic motion is also
present along this elongated direction when seen in the movie of the
image sequence.

\begin{figure}
\resizebox{\hsize}{!}{\includegraphics{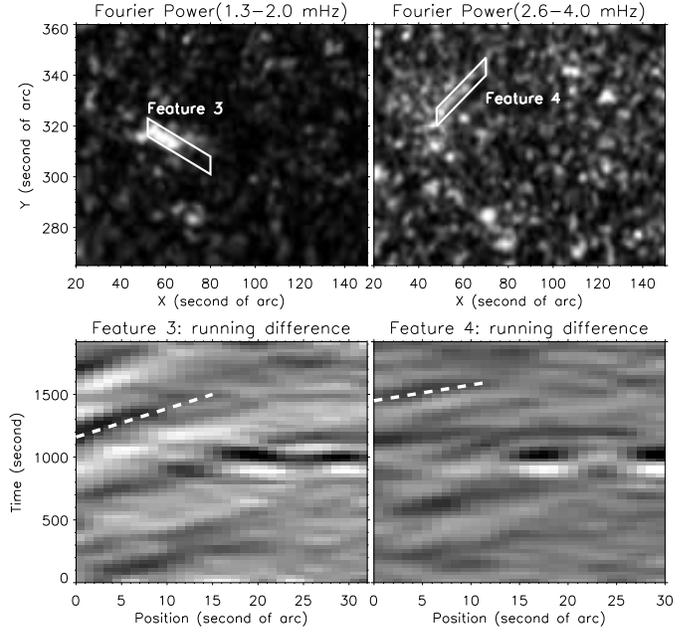}}
\caption{~Upper: the same as Fig.~\ref{fig.2}F and
Fig.~\ref{fig.2}G. The bars indicate the propagating directions of
the oscillations. Lower: plots of running difference along the long
sides of the bars shown in the upper panels. The dashed lines are
used to calculate the speed of propagating oscillations.}
\label{fig.8}
\end{figure}

The propagation speeds can be estimated if we plot the running
difference along the long side of the bars shown in the upper panels
of Fig.~\ref{fig.8}, which indicate the directions of the
propagating 10-min oscillation in feature 3 and 5-min oscillation in
feature 4 in the 171~{\AA} passband. These plots are demonstrated in
the low panels of Fig.~\ref{fig.8}. The slope of the dashed line
provides an estimate of the propagating speed, which is about
32~km/s for the 10-min oscillation, and 58~km/s for the 5-min
oscillation. These values are lower than the propagating speed of
the slow magneto-acoustic waves derived by \cite{DeMoortelEtal2000}
and \cite{DeForestGurman1998}, which are approximately 70-165~km/s
at a bright loop-footpoint and 75-150~km/s in polar plumes,
respectively. However, taking into account the line-of-sight effect
of the measured propagating speed, our values are still of the order
of the coronal sound speed, which is about 150~km/s
\citep{DeMoortelEtal2000}. Since $Alfv\acute{e}n$ oscillations are
essentially velocity perturbations and do not reveal themselves in
intensity fluctuations, our periodic propagating signatures are most
likely to be slow magneto-acoustic waves.

Here we will roughly estimate the energy flux carried by these
waves. We simply take the relationship in \cite{SakuraiEtal2002}:
\begin{equation}
\emph{$\frac{\delta I}{I}=2\frac{\rho^\prime}{\rho}=2\frac{\delta
v}{C_s}$}\label{equation1},
\end{equation}
where $\delta I/I$, $\rho^\prime/\rho$, $\delta v$ and $C_s$ are
intensity fluctuation, density fluctuation, wave velocity amplitude,
and sound speed, respectively. In our data, we found a value of
about 0.015 for $\delta I/I$. Given the sound speed of 150~km/s at
the formation temperature of the 171~{\AA} passband, we can
calculate the wave velocity amplitude as 1.05~km/s. Following
\cite{OfmanEtal1999}, the energy flux carried by slow
magneto-acoustic waves can be calculated in the following way:
\begin{equation}
\emph{$F=0.5\rho(\delta v)^2C_s$}\label{equation2},
\end{equation}
where $F$ and $\rho$ represent energy flux and density,
respectively. By using $\rho=5\times10^{-16}\rm~g~cm^{-3}$
\citep{DeMoortelEtal2000} and substituting the values of $\rho$ and
$\delta v$, we obtained an energy flux of about $\rm40~erg~
cm^{-2}~s^{-1}$ for the propagating oscillations. These values are
much lower than the required heat input to the quiet corona, which
is about $\rm3\times10^5~erg~ cm^{-2}~s^{-1}$ in
\cite{WithbroeNoyes1977}. Thus, these waves are not appropriate
candidates for coronal heating.

For the chromospheric oscillation, if we take the value of 0.04 for
$\delta I/I$ and the sound speed of 16~km/s at the formation
temperature of the 1600~{\AA} passband, we can calculate the wave
velocity amplitude as 0.32~km/s by using Equation~\ref{equation1}.
Given a particle number density of $10^{21}~cm^{-3}$ at the
temperature minimum \citep{Peter2004}, we can calculate the proton
mass density $\rho$. By substituting the values of $\rho$, $\delta
v$, and $C_s$ in the chromosphere to Equation~\ref{equation2}, we
obtain a value of $\rm1.368\times10^6~erg~ cm^{-2}~s^{-1}$ for the
energy flux carried by the waves at the emission height of the
1600~{\AA} passband. This value is very close to that estimated by
\cite{JefferiesEtal2006} and is almost one third of the required
energy budget for the chromosphere. Thus, our results support the
conclusion in \cite{JefferiesEtal2006} that low-frequency
magnetoacoustic waves provide a significant source of energy for
balancing the radiative losses of the ambient solar chromosphere.

Finally, we have to point out that the relatively short length (low
amount) of data restricts our discussion and lowers the solidness of
the results. In order to fully understand network oscillations at
boundaries of equatorial coronal holes, deep investigation to more
data sets is required in the future.

\section{Summary and conclusion}
With the help of Fourier and wavelet analyses, we studied intensity
oscillations observed simultaneously in the quiet chromosphere and
corona, above an enhanced network area at the boundary of an
equatorial coronal hole.

Images of Fourier power reveal that, oscillations above the magnetic
network show a lack of power at high frequencies (5.0-8.3 mHz), and
a significant power at low (1.3-2.0 mHz) and intermediate
frequencies (2.6-4.0 mHz) in both the 171~{\AA} and 1600~{\AA}
passbands. The former result suggests that $^{\prime\prime}$magnetic
shadows$^{\prime\prime}$ not only exist in the chromosphere, but
also extend into the lower corona. The latter result supports the
concept of $^{\prime\prime}$magneto-acoustic
portals$^{\prime\prime}$ within the network through which
low-frequency photospheric oscillations can propagate into the
chromosphere and corona.

We also studied 4 interesting features in more detail. The global
5-min oscillation is clearly present in all of the 4 analyzed
features when seen in the 1600~{\AA} passband, and is also found
with enhanced power in feature 1 (leg of a large coronal loop) and
feature 2 (legs of a coronal bright point loop) when seen in the
171~{\AA} passband. Two features above an enhanced network element
(feature 3 and feature 4) show repeated propagating behaviors with a
dominant period of 10 minutes and 5 minutes, respectively. The
derived values of the propagating speed are of the order of the
coronal sound speed. These velocities, together with the
compressional nature of the oscillation, suggest that our periodic
propagating signatures are most likely to be slow magneto-acoustic
waves.

We calculated the energy flux carried by these waves and found a
value of about $\rm40~erg~ cm^{-2}~s^{-1}$ for the 171~{\AA}
passband, which is only a small fraction of the total energy
required to heat the quiet corona. However, for the 1600~{\AA}
passband, the energy flux is about $\rm1.368\times10^6~erg~
cm^{-2}~s^{-1}$, which is of the order of the required energy budget
for the chromosphere.

\begin{acknowledgements}
The TRACE satellite is a NASA Small Explorer that images the solar
photosphere, transition region and corona with unprecedented spatial
resolution and temporal continuity. We thank C. Torrence and G. P.
Compo for providing the Wavelet software, which is available at URL:
http://paos.colorado.edu/research/wavelets/. We also thank Dr. Zhi
Xu for the helpful discussion on the related topics and the
anonymous referee for his/her careful reading of the paper and for
the comments and suggestions.

The work of H. Tian$'$s team at PKU is supported by the National
Natural Science Foundation of China(NSFC) under contracts 40574078
and 40436015. H. Tian is now supported by China Scholarship Council
for his stay in Germany.  L.-D. Xia is supported by NSFC under Grant
40574064 and the Programme for New Century Excellent Talents in
University (NCET).

\end{acknowledgements}


\begin{thebibliography}{}
    \bibitem[Banerjee et al.(2000)]{BanerjeeEtal2000}
        Banerjee, D., O$^{\prime}$Shea, E., \& Doyle, J. G. 2000, Sol.
        Phys., 196, 63
    \bibitem[Brynildsen et al.(1999)]{BrynildsenEtal1999}
        Brynildsen, N., Kjeldseth-Moe, K., Maltby, P., \& Wilhelm, K. 1999, ApJ, 517, L159
    \bibitem[Cauzzi et al.(2000)]{CauzziEtal2000}
        Cauzzi, G., Falchi, A., \& Falciani, R. 2000, A\&A, 357, 1093
    \bibitem[Curdt and Heinzel(1998)]{CurdtHeinzel1998}
        Curdt, W., \& Heinzel, P. 1998, ApJ, 503, L95
    \bibitem[Dam\'{e} et al.(1984)]{DameEtal1984}
        Dam\'{e}, L., Gouttebroze, P., \& Malherbe, J.-M. 1984, A\&A, 130, 331
    \bibitem[DeForest and Gurman(1998)]{DeForestGurman1998}
        DeForest, C. E., \& Gurman, J. B. 1998, ApJ, 501, L217
    \bibitem[De Moortel(2000)]{DeMoortelEtal2000}
        De Moortel, I., Ireland, J., \& Walsh, R. W. 2000, A\&A, 355, L23
    \bibitem[De Moortel(2002)]{DeMoortelEtal2002}
        De Moortel, I., Ireland, J., Hood, A. W., \& Walsh, R. W. 2002, A\&A,
        387, L13
    \bibitem[De Pontieu(2003)]{DePontieuEtal2003}
        De Pontieu, B., Erd\'{e}lyi, R., \& De Wijn, A. G. 2003, ApJ, 595, L63
    \bibitem[De Pontieu(2005)]{DePontieuEtal2005}
        De Pontieu, B., Erd\'{e}lyi, R., \& De Moortel, I. D. 2005, ApJ, 624,
        L61
    \bibitem[De Pontieu(2004)]{DePontieuEtal2004}
        De Pontieu, B., Erd\'{e}lyi, R., \& James, S. P. 2004, Nature, 430, 536
    \bibitem[Dowdy et al.(1986)]{DowdyEtal1986}
        Dowdy, J. F. Jr., Rabin, D., \& Moore, R. L. 1986, Sol. Phys., 105, 35
    \bibitem[Doyle(2006)]{DoyleEtal2006}
        Doyle, J. G., Popescu, M. D., \& Taroyan, Y. 2006, A\&A, 446, 327
    \bibitem[Fossum and Carlsson(2005)]{FossumCarlsson2005}
        Fossum, A., and Carlsson, M. 2005, Nature, 435, 16
    \bibitem[Fludra(1999)]{Fludra1999}
        Fludra, A. 1999, A\&A, 344, L75.
    \bibitem[Gabriel(1976)]{Gabriel1976}
        Gabriel, A. H. 1976, Philos. Trans. R. Soc. London A, 28l, 575
    \bibitem[Hansteen et al.(2006)]{HansteenEtal2006}
        Hansteen, V. H., De Pontieu, B., Rouppe van der Voort, L., van
        Noort, M., \& Carlsson, M. 2006, ApJ, 647, L73
    \bibitem[Jefferies et al.(2006)]{JefferiesEtal2006}
        Jefferies, S. M., McIntosh, S. W., Armstrong, J. D., et al. 2006,
        ApJ, 648, L151
    \bibitem[Kariyappa and Varghese(2008)]{KrijgerEtal2008}
        Kariyappa, R. and Varghese, B. A. 2008, A\&A, 485, 289
    \bibitem[Krijger et al.(2001)]{KrijgerEtal2001}
        Krijger, J. M., Rutten, R. J., Lites, B. W., et al. 2001, A\&A, 379, 1052
    \bibitem[Lites et al.(1993)]{LitesEtal1993}
        Lites, B. W., Rutten, R. J., \& Kalkofen, W. 1993, ApJ, 414, 345
    \bibitem[McAteer et al.(2002)]{McAteerEtal2002}
        McAteer, R. T. J., Gallagher, P. T.,Williams, D. R., et al. 2002,
        ApJ, 567, L165
    \bibitem[McAteer et al.(2004)]{McAteerEtal2004}
        McAteer, R. T. J., Gallagher, P. T., Bloomfield, D. S., et
        al. 2004, ApJ, 602, 436
    \bibitem[McIntosh and Judge(2001)]{McIntoshJudge2001}
        McIntosh, S. W., \& Judge, P. G. 2001, ApJ, 561, 420
    \bibitem[McIntosh and Jefferies(2006)]{McIntoshJefferies2006}
        McIntosh, S. W., \& Jefferies, S. M. 2006, ApJ, 647, L77
    \bibitem[Ofman et al.(1999)]{OfmanEtal1999}
        Ofman, L., Nakariakov, V. M., \& DeForest, C. E. 1999, ApJ, 514, 441
    \bibitem[O$^{\prime}$Shea et al.(2007)]{OSheaEtal2007}
        O$^{\prime}$Shea, E., Banerjee, D., \& Doyle, J. G. 2007, A\&A, 463, 713
    \bibitem[Patsourakos et al.(1999)]{PatsourakosEtal1999}
        Patsourakos, S., Vial, J.-C., Gabriel, A. H., \& Bellamine, N. 1999, ApJ, 522, 540
    \bibitem[Peter(2001)]{Peter2001}
        Peter, H. 2001, A\&A, 374,1108
    \bibitem[Peter(2004)]{Peter2004}
        Peter, H. 2004, Reviews in Modern Astronomy, 17, 87
    \bibitem[Popescu et al.(2005)]{PopescuEtal2005}
        Popescu, M. D., Banerjee, D., O$^{\prime}$Shea, Doyle, J. G., \&
        Xia, L. D. 2005, A\&A, 442, 1087
    \bibitem[Sakurai et al.(2002)]{SakuraiEtal2002}
        Sakurai, T., Ichimoto, K., Raju, K.P., \& Singh, J. 2002, Sol.
        Phys., 209, 265
    \bibitem[Srivastava et al.(2008)]{SrivastavaEtal2008}
        Srivastava, A. K., Kuridze, D., Zaqarashvili, T. V.,\&
        Dwivedi, B. N. 2008, A\&A, 481, L95
    \bibitem[Tian et al.(2008a)]{TianEtal2008a}
        Tian, H., Marsch, E., Tu, C.-Y., Xia, L.-D.,\& He, J.-S. 2008a,
        A\&A, 482, 267
    \bibitem[Tian et al.(2008b)]{TianEtal2008b}
        Tian, H., Xia, L.-D.,\& Li, S. 2008b, A\&A, in press
    \bibitem[Torrence and Compo(1998)]{TorrenceCompo1998}
        Torrence, C., \& Compo, G.P., 1998. Bull. Amer. Meteor. Soc. 79, 61
    \bibitem[Ugarte-Urra et al.(2004a)]{UgarteEtal2004a}
        Ugarte-Urra, I., Doyle, J. G., Madjarska, M. S., \&
        O$^{\prime}$Shea, E. 2004a, A\&A, 418, 313
    \bibitem[Ugarte-Urra et al.(2004b)]{UgarteEtal2004b}
        Ugarte-Urra, I., Doyle, J. G., Madjarska, M. S., \& Foley, C. R.
        2004b, A\&A, 425, 1083
    \bibitem[Vecchio et al.(2007)]{VecchioEtal2007}
        Vecchio, A., Cauzzi, G., Reardon, K. P.,Janssen, K.,\& Rimmele, T.
        2007, A\&A, 461, L1
    \bibitem[Withbroe and Noyes(1977)]{WithbroeNoyes1977}
        Withbroe, G. L. and Noyes, R. W. 1977, Ann. Rev. Astron. Astrophys. 15, 363
    \bibitem[Xia et al.(2005)]{XiaEtal2005}
        Xia, L.-D., Popescu, M.D., Doyle, J.G., Giannikakis, J., 2005, A\&A, 438, 1115


\end{thebibliography}
\end{document}